\documentclass[onecolumn]{revtex4}
\usepackage{amsfonts}
\usepackage{amsmath}
\usepackage{amssymb}
\usepackage{graphicx}

\begin{document}

\title{A single atom vibration sensor}

\author{Wenxi Lai$^{1}$}\email{wxlai@pku.edu.cn}\author{Yu-Quan Ma$^{1}$}\author{Qiaoxin Li$^{2}$}

\affiliation{$^{1}$ School of Applied Science, Beijing Information Science and Technology University, Beijing 100192, China}
\affiliation{$^{2}$ School of Mathematics and Science, North China Electric Power University, Beijing, 102206, China}

\begin{abstract}
Previously in vibration sensors, optical glass plates, optical fibres, carbon nanotubes, semiconductor materials, piezoelectric materials and molecules are proved to be effective transducers for sensing vibrations. In this work, for the first time, we will propose a model of vibration sensor using single atom transport in an open optical lattice. In this apparatus, information of mechanical vibration could be transferred into shaking of optical lattice through one of a cavity mirror. Shaking lattice consequently induces Mott insulator due to quantum interference. It is found that information of vibration is encoded in the atomic current and it could be extracted by Fourier transformations. The present atomic vibration sensor has wide detection range of frequency with high precision. Our present model of sensor based on atomic system opens a new area of studying vibration sensors.
\end{abstract}

\maketitle
Wide researches are carried out on vibration sensors due to their indispensable applications in detection of civil infrastructure, seismic activity, biochemical agents, structural health, and optimum performance of machines. Coupling between electrochemical reaction and vibrational signal gives rise to low frequency sensor with frequency range around $0.1$ Hz $\sim$ $10$ Hz~\cite{Zhou}. Optical glass plates can transform state of a mechanical vibration into transmission strength of laser fields, they have detectable frequency range from tens of Hz to several kHz~\cite{Kamata,Micheletto}. Optical fibres, carbon nanotubes, semiconductor materials can be applied to design micro cantilever vibration sensors which are sensitive to frequency from several Hz to hundreds of kHz~\cite{Lu,EHFeng,Strus}, or even a few tens of MHz~\cite{Masmanidis,Liang}. Sensors based on semiconductor piezoelectric materials can detect vibration with frequency from tens of MHz to GHz~\cite{Lu2015,Lu2019}. Electron-phonon interaction assisted current in molecular junctions and spectrum of photo-electron interactions in molecules can be used to fabricate molecular vibration sensors with measurement range at the order from MHz to THz~\cite{Hihath,Ritter,Tian,Chen,Bian,Liu}.

\begin{figure}
  \includegraphics[width=8.0 cm]{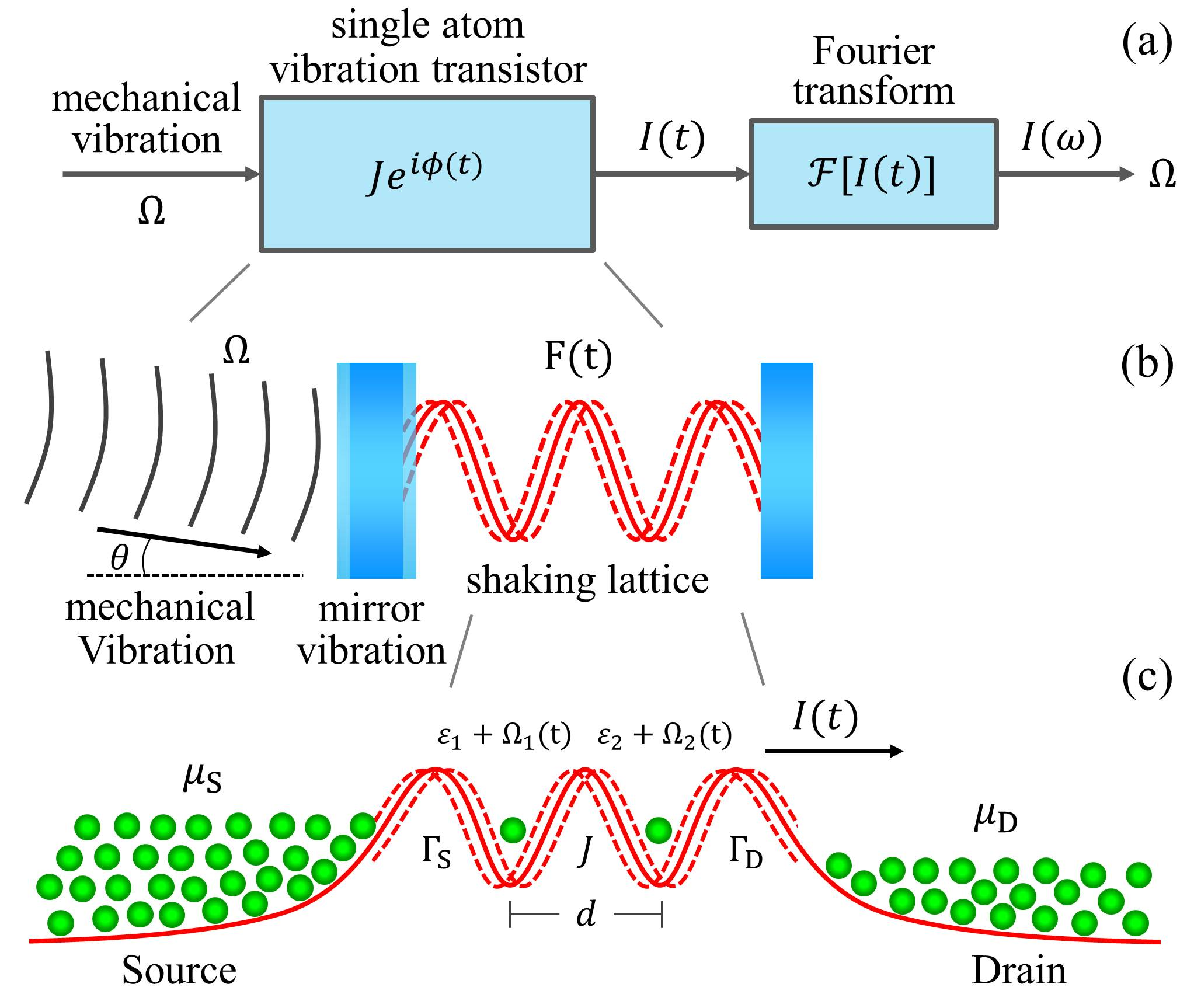}\\
  \caption{(a) Schematic structure of the single atom vibration sensor. It is composed of an single atom vibration transistor and a Fourier transformation apparatus. (2) The optomechanical coupler, in which mechanical wave of the external environment is allowed to drive the optical potential through the cavity mirror. (c) The single atom vibration transistor is an open system of cold atoms in which a double-well optical potential is coupled to two atomic reservoirs. }\label{fig1}
\end{figure}

In this paper, we propose a model of atomic vibration sensor using quantum transport of cold atoms in open optical lattices. It is inspired by phase changes in shaken optical lattices. Cold atoms in shaken optical lattices feel inertial forces which induces artificial gauge fields~\cite{Struck,Hauke,Price}. The artificial gauge fields can induce Mott insulator~\cite{Eckardt,Zenesini,Liberto}, fractional quantum Hall effect~\cite{Sorensen,Miao} and topological non-trivial states~\cite{Kitagawa,Rudner,Zheng,Wintersperger,Cheng,JYZhang}. The Mott insulator is originated from coherent destruction of tunneling~\cite{Lignier,Kierig}, depending on shaking induced Peierls phases~\cite{Eckardt2017,Sun}. Recently, sensing with quantum control of atoms and molecules is believed to be important technology for development of fundamental physics~\cite{Ye}. In the following, we will demonstrate a sensor of mechanical vibration using the phase change in a coherently controlled atomic system.

The mirror transducer used in the earlier glass based vibration sensor~\cite{Kamata,Micheletto} is an important reference case for designing our present sensor. Actually, oscillating mirrors have been applied to induce shaking of optical potentials in recent experiments~\cite{Ivanov,Zenesini}. Based on these experiments, Fig.~\ref{fig1} shows how to transfer information of an external mechanical wave into a cold atom system through a cavity mirror which can oscillate about its equilibrium position. We will prove that sensing of the mechanical wave could be achieved by taking Fourier transformations to atomic currents which is affected by the external vibration induced lattice shaking. Using a general equation of motion, the Mott insulator in atomic vibration transistor is described at arbitrary shaking frequency. The transistor is sensitive to amplitude, frequency and propagation direction of mechanical waves with high accuracy.

Center part of the atomic vibration transistor is a double-well optical potential which is coupled to Fermion atomic source and drain as articulated in Fig.~\ref{fig1}. The two wells are symmetrically arranged at positions $r_{1}=-\frac{d}{2}e_{x}$ and $r_{2}=\frac{d}{2}e_{x}$, respectively. In this model, for simplicity, we suppose atom-atom repulsion energy is too large to allow two or more atoms occupy a single lattice site. Then, the shaken open system is described by the following Hamiltonian,
\begin{eqnarray}\label{eq:Ham}
H(t) &=& \sum_{l=1}^{2}(\varepsilon_{l}+\Omega_{l}(t))n_{l}-J (a_{1}^{\dag}a_{2}+a_{2}^{\dag}a_{1})+\sum_{\alpha,k}\mu_{\alpha k}n_{\alpha k}-g\sum_{k}(a_{1}^{\dag}a_{S k}+a_{2}^{\dag}a_{D k}+H.c.),
\end{eqnarray}
where $a_{l}$ and $n_{l}=a_{l}^{\dag}a_{l}$ are annihilation and number operators of atoms at lattice site $l$ ($l=1$, $2$), respectively. $\varepsilon_{l}$ denotes bare energy level in the well $l$. In the oscillating reference, a cold atom in the well $l$ would feel a driving potential $\Omega_{l}(t)=-r_{l}\cdot F(t)$~\cite{Eckardt2017}, in which $F(t)$ is the inertial force written as $F(t)=F_{\Omega}\cos(\Omega t)(\cos(\theta)e_{x}+\sin(\theta)e_{y})$. Shaking frequency $\Omega$ of the potential is identical to frequency of the external vibration. Here, $F_{\Omega}$ is amplitude of the oscillating force and $\theta$ represents the angle between the direction of mechanical wave propagation and the one dimensional optical lattice. The second term on the right side of Hamiltonian indicates inter tunneling between the two wells with a tunneling strength $J$. In the third term, the source ($\alpha=S$) and drain ($\alpha=D$) are described by free quantum gases of single atoms with annihilation operator $a_{\alpha k}$ and number operator $n_{\alpha k}=a_{\alpha k}^{\dag}a_{\alpha k}$. Here, $\mu_{\alpha k}$ represents energy of a single atom with momentum $k$. The last term in the Hamiltonian represents coupling energy between the shaken system and atomic reservoirs (the source and drain) with the coupling amplitude $g$.

According to standard derivation of quantum master equation under Born-Markovian approximation in quantum optics~\cite{Scully,Lai}, equation of motion of the shaking open system could be deduced based on Hamiltonian \eqref{eq:Ham},
\begin{eqnarray}\label{eq:QME}
\frac{\partial}{\partial t}\rho&=&-i[\varepsilon_{1}n_{1}+\varepsilon_{2}n_{2}-(Je^{i\phi(t)}a_{1}^{\dag}a_{2}+H.c.),\rho]+\mathcal{L}_{S}\rho+\mathcal{L}_{D}\rho,
\end{eqnarray}
where $\rho$ is the system density matrix and $\phi(t)$ indicates the time dependent Peierls phase $\phi(t)=\frac{K}{\hbar\omega}\cos(\theta)\sin(\Omega t)$. The Peierls phase is proportional to the shaking strength $K=F_{\Omega}d$ in which $d$ denotes distance between the two wells as shown in Fig.~\ref{fig1}. The Lindblad super operators depict atom transport between the system and atomic electrodes, they have following typical expressions,
\begin{eqnarray}\label{eq:LSOL}
\mathcal{L}_{S}\rho&=&\frac{\Gamma}{2}(1-f_{S}(\varepsilon_{1}))(2a_{1}\rho a_{1}^{\dag}-\{a_{1}^{\dag}a_{1},\rho\})+\frac{\Gamma}{2}f_{S}(\varepsilon_{1})(2a_{1}^{\dag}\rho a_{1}-\{a_{1}a_{1}^{\dag},\rho\})\end{eqnarray}
and
\begin{eqnarray}\label{eq:LSOR}
\mathcal{L}_{D}\rho&=&\frac{\Gamma}{2}(1-f_{D}(\varepsilon_{2}))(2a_{2}\rho a_{2}^{\dag}-\{a_{2}^{\dag}a_{2},\rho\})+\frac{\Gamma}{2}f_{D}(\varepsilon_{2})(2a_{2}^{\dag}\rho a_{2}-\{a_{2}a_{2}^{\dag},\rho\}), \end{eqnarray}
where the brace $\{,\}$ represents anti commutation. The two coupling rates are set to be equal, $\Gamma=2\pi D(\varepsilon_{1})|g|^{2}=2\pi D(\varepsilon_{2})|g|^{2}$, in which the function $D$ denotes density of energy states in the source and the drain. Chemical potentials of the source $\mu_{S}$ and the drain $\mu_{D}$ are appeared in the corresponding Fermi-Dirac distribution functions $f_{S}(\varepsilon_{1})=\frac{1}{e^{(\varepsilon_{1}-\mu_{S})/k_{B}}+1}$ and $f_{D}(\varepsilon_{2})=\frac{1}{e^{(\varepsilon_{2}-\mu_{D})/k_{B}T}+1}$. Here, $k_{B}$ is the Boltzmann constant and $T$ is temperature in the two atomic reservoirs. Eq. \eqref{eq:QME}, as a differential equation of the density matrix $\rho$ with time variable coefficient, could be written in the form $\frac{\partial}{\partial t}\rho=M(t)\rho$ and it has a solution $\rho=e^{\int_{0}^{t}M(t')dt'}\rho(0)$ for a given initial state $\rho(0)$. Hilbert space of the double-well system is generated from the basic states $|00\rangle$, $|01\rangle$, $|10\rangle$ and $|11\rangle$. The given parameters throughout this article are $J/2\pi=200$ Hz~\cite{Mancini,Livi}, $\Gamma=0.1J$, $k_{B}T=0.1J$ ($T=0.96$ nK), $\varepsilon_{1}=\varepsilon_{2}=5J$, $\mu_{S}=10J$ and $\mu_{D}=0$. Additionally, $\theta=0$ except the last figure.

\begin{figure}
  \includegraphics[width=16cm]{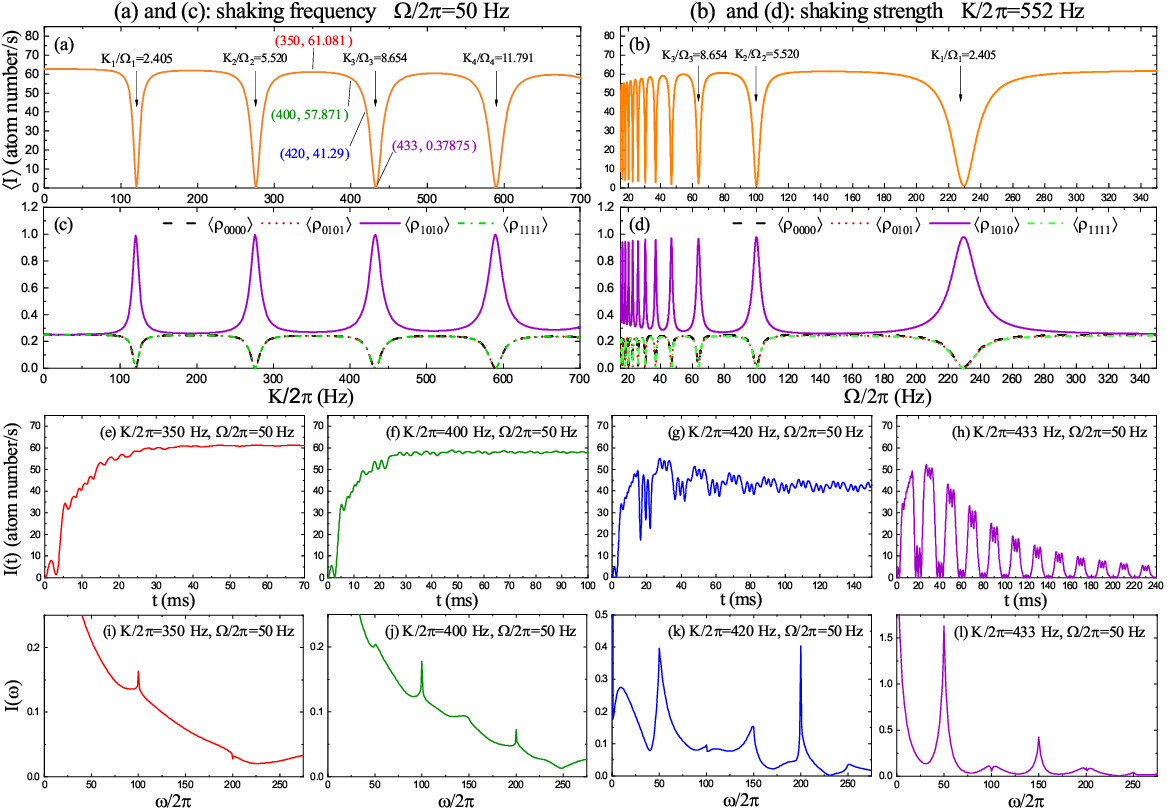}\\
  \caption{(Color on line) (a) Average current as a function of shaking amplitude $K$. (b) Average current as a function of shaking frequency $\omega$. (c) Changes of diagonal density matrix elements corresponding to the average current in (a).(d) Changes of diagonal density matrix elements corresponding to the average current in (b). (e)-(h) Current behavior versus real time under different values of shaking amplitude chosen from (a). (i)-(l) The current spectrums corresponding to the current fluctuations in (e)-(h), respectively.}\label{fig2}
\end{figure}

Time dependent current is direct outcome of the atomic vibration transistor. The current could be derived from the continuity equation $I_{S}-I_{D}=\frac{\partial}{\partial t}Tr_{S}[\rho(t)\sum_{l=1}^{2}a_{l}^{\dag}a_{l}]$~\cite{Davies}, where $I_{S}$ and $I_{D}$ are currents detected in the source and drain, respectively. $Tr_{S}$ indicates trace over the system states. Using this formula, we achieve the right side current $I(t)=-\Gamma f_{D}(\varepsilon_{2})(\rho_{00,00}+\rho_{10,10})+\Gamma(1-f_{D}(\varepsilon_{2}))(\rho_{01,01}+\rho_{11,11})$, where $I(t)=I_{D}(t)$ for simplicity. This current expression could be equivalently derived from many-body Schr\"{o}dinger equation approach~\cite{Gurvitz}. For the atomic vibration sensor, Fourier transformation of the current $I(t)$
\begin{eqnarray}\label{eq:FT}
    I(\omega)&=&\int_{0}^{\tau}I(t)e^{-i\omega t}dt
\end{eqnarray}
is significant. Here, time $\tau$ takes infinitely large value, $\tau\rightarrow\infty$, as the current $I(t)$ is generally not a periodic function. It is taken be $\tau=1$ s in all the present numerical treatments. Modulus of the Fourier transformation $|I(\omega)|=\sqrt{Re[I(\omega)]^{2}+Im[I(\omega)]^{2}}$ is taken in our results. Another important quantity for the sensor is the timely averaged current,
\begin{eqnarray}\label{eq:AC}
\langle I\rangle=\frac{1}{\tau}\int_{0}^{\tau}I(t)dt,
\end{eqnarray}
which can be seen as zero order term of the Fourier transformation with respect to $\omega=0$ in Eq.\eqref{eq:FT}.

Firstly, average currents calculated using Eq.\eqref{eq:AC} are plotted in Figs.~\ref{fig2} (a) and (b). They show there are particular points where current $\langle I\rangle$ is close to be zero. This effect appears when the ratio of vibration strength and frequency $K/\Omega$ takes certain values, such as $K_{1}/\Omega_{1}=2.405$, $K_{2}/\Omega_{2}=5.520$, $K_{3}/\Omega_{3}=8.654$, $K_{4}/\Omega_{4}=11.791$, $K_{5}/\Omega_{5}=14.931$, $K_{6}/\Omega_{6}=18.071$, $K_{7}/\Omega_{7}=21.213$, $K_{7}/\Omega_{7}=24.352$ and so on. One can check these ratios that they approximately form an arithmetic sequence with a common difference of $\pi$. They have a general formula as,
\begin{eqnarray}\label{eq:GCMI}
   \frac{K_{n}}{\Omega_{n}}-\frac{K_{2}}{\Omega_{2}}\approx(n-2)\pi,
\end{eqnarray}
where $\frac{K_{2}}{\Omega_{2}}=5.520$ and $n$ is any positive integer number. The first point, $K_{1}/\Omega_{1}\approx2.4$, for obtaining Mott insulator phase has been reported in previous studies on driven optical lattices~\cite{Zenesini,Eckardt,Lignier}. Therefore, the empirical formula \eqref{eq:GCMI} is a general equation of the factor $K_{n}/\Omega_{n}$ for occurrence of Mott insulator. In our atomic vibration sensor, it can supply information of vibration frequency and strength. In Eq.\eqref{eq:GCMI}, there are three variables, $K_{n}$, $\Omega_{n}$ and $n$. One can set $K_{n}$ and $n$ by choosing system parameters to design a vibration sensor for verification of a particular frequency $\Omega_{n}$. Alternatively, one can test a particular $K_{n}$ when $n$ and $\Omega_{n}$ are known.

To understand the Mott insulator mentioned above, diagonal elements of density matrix has been shown in Figs.~\ref{fig2} (c) and (d) under the identical parameters as in Figs.~\ref{fig2} (a) and (b). Both of these two figures illustrate the current destructive occurs at the point $K_{n}/\Omega_{n}$ when atom population is localization in one of the two optical wells. It is originated from the coherent destruction of tunneling in shaking optical lattices due to destructive interference in coherent tunnelings~\cite{Kierig,Lignier}. As a result, atom transition from one site to the other site would be blocked. It is similar to coherent population trapping in quantum optics where atoms are coherently trapped in one of its internal electronic states~\cite{Whitley,Arimondo}. The later effect is also explained by quantum interference~\cite{Scully}.

\begin{figure}
  \includegraphics[width=16.0cm]{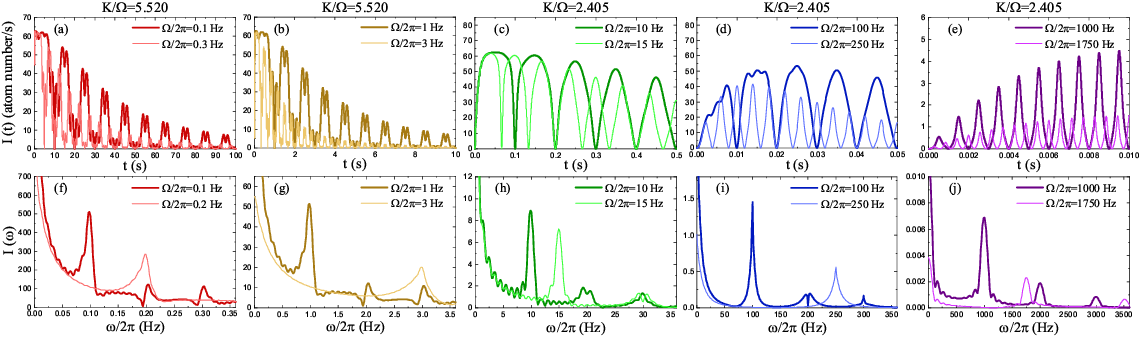}\\
  \caption{(Color on line) (a)-(e) Current as function of real time at different shaking frequency $\Omega$. (f)-(j) Current spectrums corresponding to the current fluctuations in (a)-(e), respectively.}\label{fig3}
\end{figure}

In fact, the vibration sensor can directly detect vibrational frequency. To show this process we choose four typical points in Fig.~\ref{fig2} (a), corresponding to four different shaking strengths $K/2\pi=350$ Hz, $400$ Hz, $420$ Hz and $433$ Hz, but the same shaking frequency $\Omega/2\pi=50$ Hz. Time evolution of atomic currents $I(t)$ are shown in Figs.~\ref{fig2} (e)-(h) with respect to these four shaking amplitudes. Consequently, frequency spectrum of these time dependent currents are obtained through the Fourier transformation given in Eq.\eqref{eq:FT}. Spectrums of currents in Figs.~\ref{fig2} (i)-(l) reveal frequency of shaking potential is encoded in atomic currents of the transistor. In other words, frequency of the vibration could be directly extracted from Fourier transformations of atomic currents. For example, at the point $K/2\pi=350$ Hz in Fig.~\ref{fig2} (a), corresponding spectrum of current plotted in Fig.~\ref{fig2} (i) supplies information of the frequency as $\omega=2\Omega$ and $4\Omega$. Fortunately, in the area represented by the points $K/2\pi=400$ Hz, $420$ Hz and $433$ Hz, the exact frequency $\omega=\Omega$ could be read as illustrated in Figs.~\ref{fig2} (j)-(l).

Current destruction effect is significant for direct read of the vibrational frequency. The three points $K/2\pi=400$ Hz, $420$ Hz and $433$ Hz in Fig.~\ref{fig2} (a) are located in the area of current destruction which is originated from the time dependent Peierls phase $\phi(t)$. The time dependent Peierls phase $\phi(t)$ leads to polarization of density matrix, as shown in Figs.~\ref{fig2} (c) and (d), through the evolution process $\rho(t)=exp[i\int_{0}^{\tau}M(t)dt]\rho(0)$. Therefore, the current destructive is originated from the Peierls phase. Furthermore, the Peierls phase $\phi(t)$ directly comes from the mechanical vibration induced inertial potential $\phi(t)=\int_{0}^{t}\Omega(t)dt'$. It indicates information of the mechanical vibration is transferred to atomic current through the time dependent Peierls phase.

\begin{figure}
  \includegraphics[width=8.6cm]{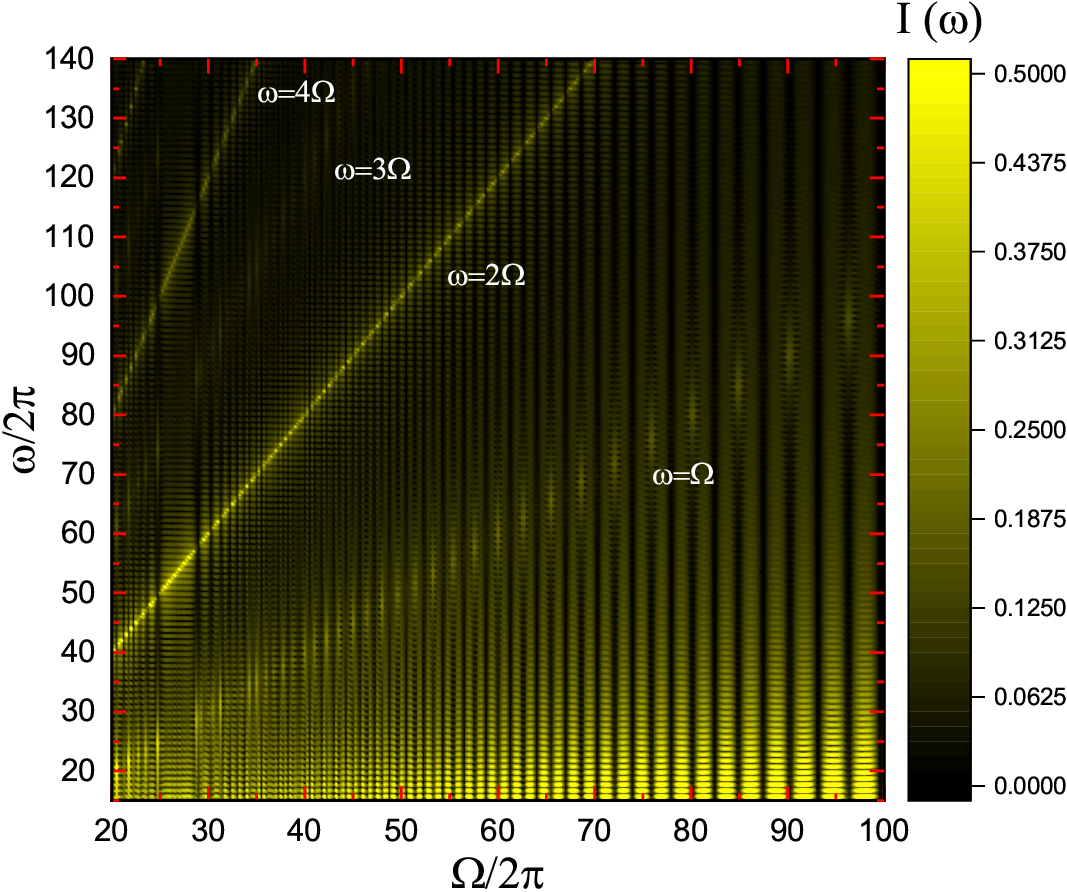}\\
  \caption{(Color on line) Frequency spectrum of current with respect to shaking frequency $\Omega$ under the given shaking amplitude $K=45J$.}\label{fig4}
\end{figure}

The present sensor has large detection range of frequency. Fig.~\ref{fig3} shows frequency monitoring from the value $0.1$ Hz to $1$ kHz. In fact, frequency lower than $0.1$ Hz and higher than $1$ kHz could be detected as soon as the vibration frequency $\Omega$ and vibration strength $K$ satisfy Eq.\eqref{eq:GCMI}. With contemporary cold atom experiments, the sensor may encounter some challenges facing ultra low or ultra high vibrational frequency. For much low frequency, for example, $\Omega=0.1$ Hz as shown in Figs.~\ref{fig3} (a) and (f), one needs several tens of seconds to monitor atomic current. For much higher frequency, for example, $\Omega=1000$ Hz as shown in Figs.~\ref{fig3} (e) and (j), one should require large shaking amplitude around $K=2.405\Omega$ to read clear information of the frequency.

Frequency spectrum of current with respect to continuous change of vibrational frequency $\Omega$ with a definite shaking strength $K=45 J$ is shown in Fig.~\ref{fig4}. Peaks of the spectrums are appeared at the positions $\omega=n\Omega$ ($n$ is positive integer number). It is clear that frequency $\Omega$ of the mechanical vibration could be found directly from Fourier transformation $I(\omega)$ of the current. In this figure, one observes many (black) gaps where the expected vibrational frequency could not be observed. Gaps in the spectrum for a given shaking strength $K$ could be supplemented by tuning $K$ through adjusting the lattice constant $d$, since $K=F_{\Omega}d$.

\begin{figure}
  \includegraphics[width=8.5cm]{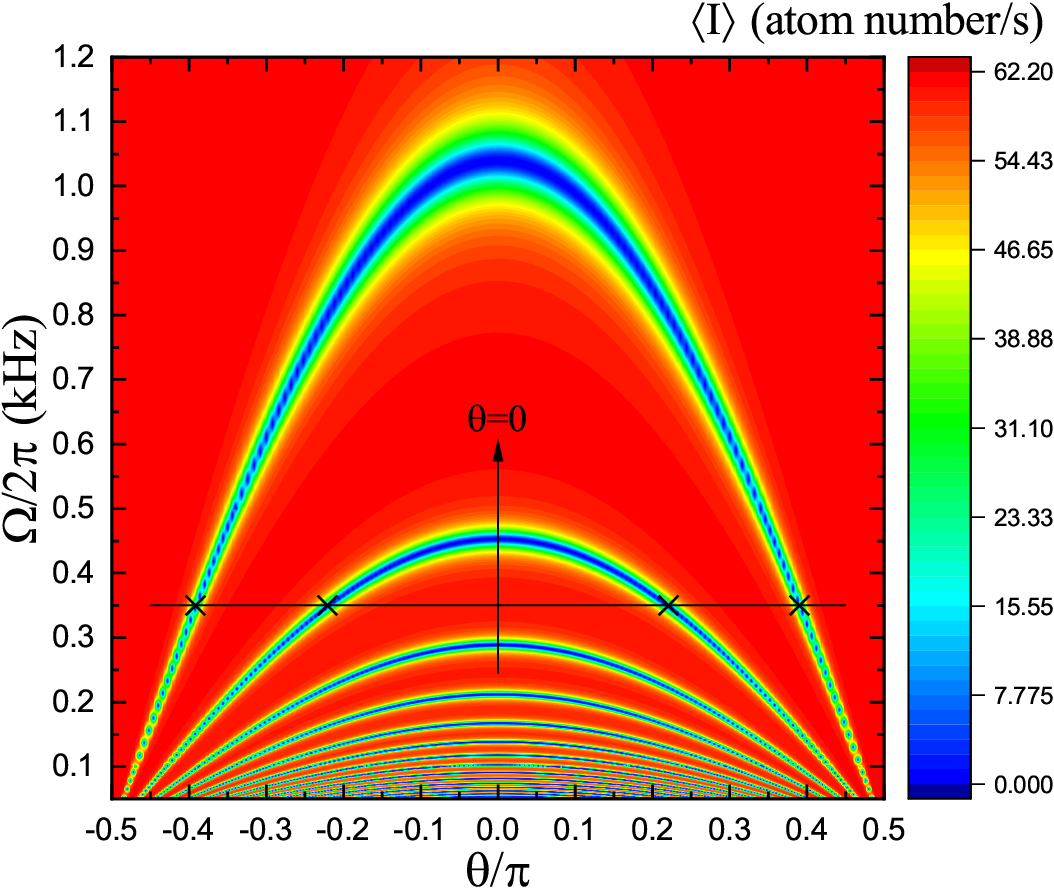}\\
  \caption{Average current as a function of shaking frequency $\Omega$ and mechanical wave direction $\theta$ with respect to the given shaking amplitude $K=12.5J$.}\label{fig5}
\end{figure}

There is a simple method to judge direction of a mechanical wave. This method takes the advantage of Peierls phase $\phi(t)$ which is proportional to $K\cos(\theta)$. Here, $\theta$ is angle between direction of the atomic vibration sensor and the mechanical wave to be measured as illustrated in Fig.~\ref{fig1}. Therefore, changing direction of the vibrational sensor is equivalently changing the shaking strength. For a vibration with a given frequency $\Omega$ originated from a monochromatic mechanical wave, there are even number of current destruction (Mott insulator) would occur by tuning $\theta$ from $-\pi/2$ to $\pi/2$. This results have been shown in Fig.~\ref{fig5} for different frequency $\Omega$. For example, if the frequency of vibration is $\Omega=0.35$ kHz, current destruction should be observed four times. The even number positions of the current destruction are symmetrically arranged on two side of the line in the direction $\theta=0$. Then, one can point out direction of the mechanical wave propagation is parallel to the line on angle $\theta=0$. Generally, the direction could point to any angle, depending on the choice of coordinate system.

In experiments, the oscillating mirror could works as a transducer which transfer an external mechanical wave into shaking of optical potential in a cavity~\cite{Ivanov,Zenesini}. It can be comprehended from periodically modulated distance between two mirrors of a cavity. The modulated distance between the two mirrors would change frequency of laser field which generate standing wave in the cavity. Furthermore, frequency change of laser beams lead to shaking of optical lattices~\cite{Struck,Reitter}. For progresses in open optical systems, some simply structured open optical potentials with non-equilibrium atomic gases have been experimentally implemented recent years~\cite{Caliga17,Krinner,Amico}. Corresponding theoretical works on atomic transport in open optical lattices are studied for future applications of atomtronic transistors and devices~\cite{Seaman,Caliga,Wilsmann}.

As conclusions, a vibration sensor based on single atom quantum interference in an open optical lattice has been proved here. Peierls phase induced quantum interference is the main reason of the coherent destruction of tunneling and Mott insulator. Due to the quantum interference, near the point of Mott insulator, atomic current fluctuates and decreases remarkably. The Mott insulator results from vibration of the optical lattice, in which the vibration is supposed to be transferred from external vibration to be tested. With numerical calculation of the quantum master equation, we find vibration frequency and Vibration strength satisfy a general formula which leads to Mott insulator in the single atom vibration transistor. Fourier transformations of the fluctuating currents can be used to read accurate information of vibrations. This is principle of the single atom vibration sensor. Quantum evolution of the system is described by time variable coefficient quantum master equation. To obtain the results here only Born-Markovian approximation is considered during the theoretical treatment.

\begin{flushleft}
  \textbf{AUTHOR DECLARATIONS}
\end{flushleft}

\begin{flushleft}
\textbf{Conflict of Interest}: The authors have no conflicts to disclose.
\end{flushleft}

\begin{flushleft}
\textbf{Author Contributions}: \textbf{Wenxi Lai:} Conceptualization (lead); Data curation(equal); Formal analysis (lead); Funding acquisition (equal); Investigation (equal); Methodology (lead); Project administration (equal); Resources (lead); Software (equal); Supervision (lead); Validation (lead); Visualization (equal); Writing-original draft (lead); Writing-review \& editing (equal). \textbf{Yu-Quan Ma:} Formal analysis (supporting); Funding acquisition(equal); Investigation (equal); Methodology (supporting); Project administration (equal). \textbf{Qiaoxin Li:} Data curation (equal); Software (equal); Validation (supporting); Visualization (equal); Writing-review \& editing (equal).
\end{flushleft}

\begin{flushleft}
\textbf{Data Availability Statement}: The data that supports the findings of this study are available within the article.
\end{flushleft}

\begin{acknowledgments}
This work was supported by R \& D Program of Beijing Municipal Education Commission (BMEC) (Grant No. KM202011232017) and Natural Science Foundation of Beijing Municipality (Grant No. 1232026).
\end{acknowledgments}

\end{document}